\documentclass[referee]{aa} 
\usepackage{graphicx}
\begin{document}
   \title{Influence of stellar X--ray luminosity distribution and its evolution on exoplanetary mass loss}

   \author{T. Penz
          \inst{1}
          \and
          G. Micela\inst{1}
          \and
          H. Lammer\inst{2}
          }

   \offprints{T. Penz}

   \institute{INAF--Osservatorio Astronomico di Palermo, Piazza del Parlamento 1, I--90134 Palermo, Italy\\
              \email{tpenz@astropa.inaf.it; giusi@astropa.inaf.it}
         \and
             Space Research Institute, Austrian Academy of Sciences, Schmiedlstr. 6, A-8042 Graz, Austria\\
             \email{helmut.lammer@oeaw.ac.at}
                          }

   \date{Received ; accepted }

  \abstract{}
   {We investigate the influence of high--energy
   stellar radiation at close--in orbits on atmospheric mass loss during stellar evolution of a G--type star.}
   {High--energy stellar luminosity varies over a wide range for G field stars.
   The temporal evolution of the distribution of stellar
   X--ray luminosity and its influence on the evolution of
   close--in exoplanets is investigated. X--ray luminosity distributions from the Pleiades, the Hyades and the field
   are used to derive a scaling law for the evolution of the
   stellar X--ray luminosity distribution. A modified
   energy--limited escape approach is used to calculate
   atmospheric mass loss for a broad range of planetary parameters.}
   {We show that the evolution of close--in exoplanets strongly
   depends on the detailed X--ray luminosity history of their host stars, which varies
   over several orders--of--magnitude for G stars. Stars located at the
   high--energy tail of the luminosity distribution can evaporate
   most of its planets within 0.5 AU, while for a moderate luminosity a
   significant fraction of planets can survive. We show the
   change on an initial planetary mass distribution caused
   by atmospheric escape.}
   {}

   \maketitle
%

\section{Introduction}

   Since the discovery of the first close--in Hot Jupiter in 1995
   (Mayor and Queloz, 1995) the investigation of the
   structure and the temporal evolution of their highly irradiated
   atmospheres was a main topic for both, modelers and observers.
   In 2002, Charbonneau et al. (2002) reported the
   first detection of an atmosphere around HD209458b by observing
   absorption in the NaI D lines during transits. Soon after
   that Vidal-Madjar et al. (2003, 2004) reported the detection of several
   other species and pointed out that the planet is losing mass at
   a rate of more than 10$^{10}$ g/s, as indicated by an expanded
   atmosphere likely due to heating by incoming stellar radiation.
   Recently, a layer of excited hydrogen atoms with a
   temperature of about 5000 K slightly above the visual radius of
   the planet was reported (Ballester et al., 2007), further
   strengthening the hypothesis of a strongly heated atmosphere.
   Shortly after the discovery of Hot Jupiters, theoretical models
   were established to verify the evaporation conditions of these
   close--in planets (Guillot et al., 1996). Lammer et al. (2003)
   used an energy--limited escape approach for investigating the
   atmospheric escape from close--in exoplanets. After that,
   several hydrodynamic models were established in order to
   analyze the atmospheric conditions in more detail (Yelle, 2004,
   2006; Tian et al., 2005; Garcia Munoz, 2007). Cecchi--Pestellini et al. (2006) developed
   an accurate heat transfer model to determine the heating of
   planetary atmospheres due to X--rays. Mass loss
   calculations were also included in different models for giant
   planet evolution (Lecavelier des Etangs et al., 2004; Baraffe et
   al., 2004, 2005, Hubbard et al., 2006, 2007). Recently, Lecavelier des Etangs (2007)
   presented a diagram representation where the influence of
   atmospheric evaporation on the evolution of the exoplanets can
   be estimated for a wide range of planetary parameters and for
   different spectral types of host stars.\\
   However, in all these works the present--day solar luminosity
   was scaled to the considered orbital distances, or, in the
   evolutionary models, the scaling law derived by Guinan and
   Ribas (2002) from the \textit{Sun in Time} program was used.
   This program was based on a small sample of solar--type stars,
   allowing to determine an average scaling law for the temporal evolution
   of the stellar radiation for a range of wavelength from
   1--1200 \AA. However, in reality, G stars show a broad
   distribution of the luminosity which varies over a few
   orders--of--magnitude. This distribution can be observed only
   for the very extreme UV and X--ray (1--200 \AA) (e.g., Preibisch
   and Feigelson, 2005) because of interstellar absorption and lack
   of sensitive instruments. In this paper, we will focus on the influence
   of radiation in this wavelength range, which is justified since
   Cecchi--Pestellini et al. (2006) showed that X--ray and very
   extreme UV radiation has significant influence on planetary
   atmospheres. Also the X--ray luminosity of clusters with a given age like the
   Pleiades (Micela, 2001) and the Hyades (Stern et al., 1995) is known,
   therefore the temporal evolution of the population of G stars can be
   derived. Using this X--ray luminosity distribution as an input,
   we apply a modified energy--limited approach formula (Erkaev et
   al., 2007) to study the influence of atmospheric loss for
   close--in exoplanets around solar--like G--stars and discuss the expected deviations from
   an initially known planetary mass distribution due to
   atmospheric loss.


\section{Stellar X--ray luminosity distribution and temporal evolution of the radiation flux}

\begin{figure}
   \centering
   \includegraphics[width=9cm]{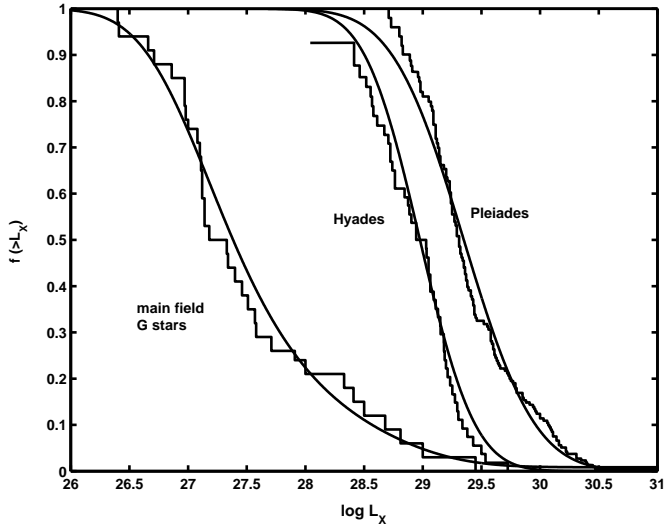}
      \caption{Comparison between the observed and approximated
      cumulative X--ray
      distribution function for G stars from the main field (Schmitt, 1997),
      the Hyades (Stern et al., 1995), and the Pleiades (Micela, 2002).}
         \label{comp}
\end{figure}

\begin{figure}
   \centering
   \includegraphics[width=9cm]{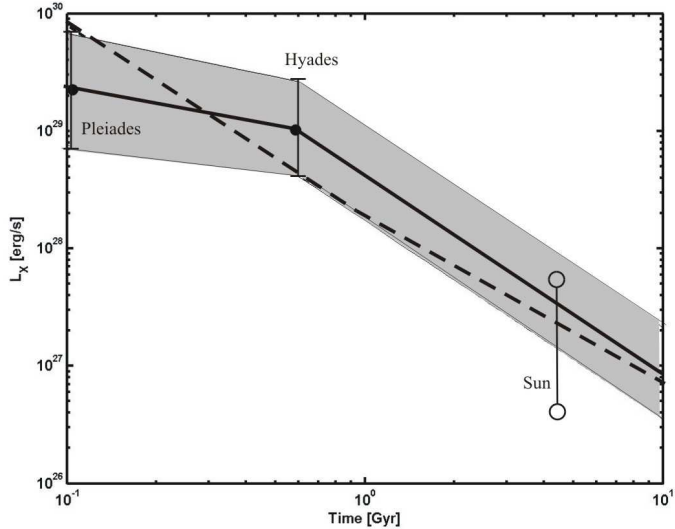}
      \caption{L$_{X}$ from Eq. \ref{scaling} compared with the
      scaling by Ribas et al. (2005) for 1--100 $\mbox{\AA}$ (dashed line).
      The shaded area gives the 1$\sigma$
      equivalent derived from the log--normal distribution for the Hyades and the
      Pleiades. For comparison we show L$_{X}$ of the Sun for maximum and minimum
      activity (empty circles) from Peres et al. (2000).}
         \label{lumi}
   \end{figure}

The \textit{Sun in Time} program was established to trace the
spectral irradiance of the Sun over its lifetime (Dorren and
Guinan, 1994). Thus, a sample of more than 15 accurately selected
solar proxies with different age was studied. To determine the
high--energy emissions this sample was reduced to 6 stars which
were observed by different high--energy instruments (Ribas et al.,
2005). Therefore, it is obvious that this study is limited to a
small sample of accurately chosen stars, but exoplanets are
orbiting a much broader range of stars, which makes it necessary
to study the whole range of possible luminosities. Indeed, one of
the most interesting and unexpected results of the first imaging
X--ray telescopes was that stars, and in particular solar--type
stars, show a broad range of emission (Vaiana et al., 1981). This
broad range is mainly due to the age--dependence of X--ray
luminosity, but it has been shown that even in samples with the
same age, as in open clusters, a spread of at least one
order--of--magnitude is present (e.g., Stern et al., 1995; Micela
et al., 1996). The observed spread in $L_x$ at fixed age is
associated to the spread in rotational periods, to which the level
of activity is linked (Pizzolato et al., 2003), that depends on
circumstellar disk evolution in the pre--main sequence phase. In
light of the discussion above, it is needed to consider the
evolution of the complete luminosity distribution in order to
understand the possible effects on planetary atmospheres. In this
study we will focus only on the luminosity distribution of G--type
stars. Because of interstellar absorption, it is not possible with
present instrumentation to achieve information about the
luminosity of a large sample of stars in the EUV (200--900
$\mbox{\AA}$), so in this work we are focussing on X--ray data.
While the entire band should be considered, X--rays have however
an important role inducing two classes of effects. First, they
modify the ionization and the chemical equilibrium of the outer
planetary atmosphere, and second, they produce a significant
population of secondary electrons that can penetrate down in the
atmosphere contributing to its heating
(Cecchi--Pestellini et al., 2006).\\
To get information about the temporal evolution of the X--ray
luminosity we constructed a scaling law using data of stellar
clusters with a known age and data from the a sample of nearby
solar--type stars. For a given cluster, we parameterize the
logarithm of the X--ray luminosity distribution function following
a log--normal distribution and then calculate the distribution for
the whole sample observed in the solar neighborhood where a mix of
age is present (Schmitt, 1997). The cumulative distribution
function for a log--normal distribution is given as
\begin{equation}\mbox{CDF}(ln(L_{X}))=\frac{1}{2}+\frac{1}{2}\mbox{erf}[\frac{ln(L_{X})-\mu}{\sigma
\sqrt{2}}]\,,\end{equation} where $\mu$ and $\sigma$ are the mean
and standard deviation of the variable's logarithm. As a
representative of young stars we used the Pleiades cluster, which
has an estimated age of 100 Myr (Stauffer et al., 2005, and
references therein). The maximum likelyhood luminosity function of
G stars in the Pleiades clusters is given in Micela (2002). This
can be fitted well with a log--normal distribution with
$\mu=67.58$ (corresponding to $L_X=10^{29.35}$ erg/s) and
$\sigma=1.1$ (Fig. \ref{comp}). For intermediate age stars we use
the Hyades cluster with an age of about 600 Myr (Stauffer et al.,
2005, and references therein), with a luminosity function given in
Stern et al. (1995). The log-normal parameters for the Hyades are
$\mu=66.8$ ($L_X=10^{29.0}$ erg/s) and $\sigma=0.9$ (Fig.
\ref{comp}). For 4.6 Gyr old stars we assume the same standard
deviation as for the Hyades and a mean value of $\mu=63.3$
consistent with present day solar emissions. Under the assumption
of a constant standard deviation over time, we can derive the
evolution of the total distribution using a scaling law just for
the mean value according to
\begin{equation}\label{scaling}L_{X}=\biggl\{\begin{array}{cc}0.375L_{0}t^{-0.425} & \quad t\leq 0.6\,\, \mbox{Gyr}\\0.19L_{0}t^{-1.69} & \quad t>0.6\,\, \mbox{Gyr}\end{array}\,,\label{scaling}\end{equation}
where $L_{0}$ is the mean log luminosity of the Pleiades
($10^{29.35}$ erg/s). The derived scaling law is a first
approximation limited by the clusters available \footnote{Many
other clusters have been observed in X--rays, consistent with our
scaling law, but not with an age greater than 600 Myr. For
sensitivity reasons, it has been not possible to derive the X--ray
luminosity for clusters with solar--age stars.}. In Fig.
\ref{lumi} we compare the scaling law from Eq. \ref{scaling} with
those derived by Ribas et al. (2005). The solid line represents
the mean value of the log--normal distribution, while the shaded
area is bounded by curves giving the 1$\sigma$ equivalent spread.
The dashed line gives the 1--100 $\mbox{\AA}$ interval from Ribas et al. (2005).\\
To verify the validity of the derived scaling law we construct the
distribution function for the nearby field stars by assuming a
constant stellar birth rate and a maximum age of 12 Gyr. Doing so,
we can very well reconstruct the cumulative distribution function
for G stars given in Schmitt (1997) (Fig. \ref{comp}).

\section{Atmospheric loss calculations}

To calculate the atmospheric loss from exoplanets we use a
modified energy--limited approach, which was discussed in details
by Erkaev et al. (2007), so we shall be brief here. It is based on
the work by Watson et al. (1981) who developed a model for an
escaping hydrogen atmosphere, assuming that it is bound by gravity
at the lower boundary, that the total incoming energy is absorbed
in a narrow region where the optical depth is unity, and that the
main source of heating is stellar radiation. Erkaev et al. (2007)
included the influence of Roche lobe effects, which are important
at close--in distances. The planetary mass loss is given as
\begin{equation}\frac{dM}{dt}=\frac{4\pi R_{pl}^{3}F_{X}}{mM_{pl}GK}\,,\end{equation}
where we assumed for simplicity that the radius where most of the
absorption takes place is close to the planetary surface. Here,
$R_{pl}$ and $M_{pl}$ are the radius and mass of the planet, $m$
is the mass of the hydrogen atom, $G$ is the gravitational
constant, $F_{X}$ is the flux at the planets orbit derived from
$L_{X}$, and $K$ takes into account Roche lobe effects (Erkaev et
al., 2007). We can verified the formula for the case of HD209458b.
If we take $R_{pl}=1.32$ $R_{jup}$ and $M_{pl}=0.65$ $M_{pl}$
(Knutson et al., 2007), and assume the stars luminosity to be
$1.1\times 10^{27}$ erg/s. We derive this value from an X-ray
observation made with XMM/Newton observatory. An X-ray source with
a count rate of $3.3 \cdot 10^{-3}$ cts/s is found at the stellar
position of HD 209458. Assuming a hydrogen column $ N_H =10^{19}$
cm$^{-2}$, a coronal temperature of $T=3\cdot 10^6$ K and a
distance of 47 pc we derive $L_X~=~ 1.1\cdot 10^{27}$ erg
s$^{-1}$. For these values a loss rate of $1.25\times10^{10}$ g/s
is found, which is in agreement with observational limitations
(Vidal--Madjar et al., 2003, 2004). One should note that the
derived loss rate represents a lower limit, since it is based
solely on the X--ray flux and does not consider the EUV flux from
100--900 $\mbox{\AA}$ in the energy budget. However, also the
observation by Vidal--Madjar et al. (2003, 2004) presents a lower
limit for the mass loss.

\subsection{Distributions for a single initial mass}

\begin{figure}
   \centering
   \includegraphics[width=9cm]{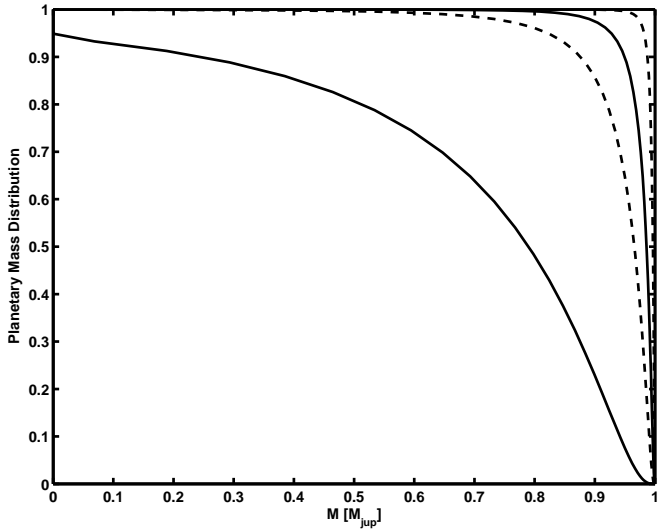}
      \caption{Planetary mass distribution after 4 Gyr for a
      initial mass of 1 M$_{jup}$, densities of 0.4 (solid lines)
      and 1.4 (dashed lines) g/cm$^{3}$, and orbital distances
      of 0.02, and 0.05 AU (starting from the lower
      curve).}
         \label{single_jup}
   \end{figure}

\begin{figure}
   \centering
   \includegraphics[width=9cm]{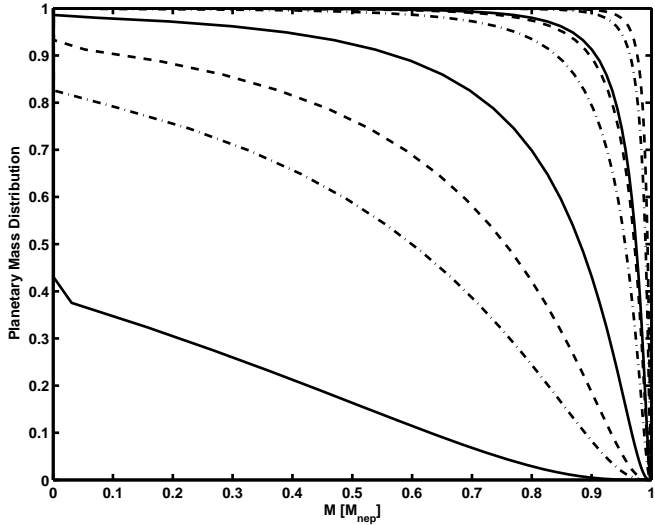}
      \caption{Similar to Fig. \ref{single_jup}, but for a
      initial mass of 1 M$_{nep}$, and densities of 0.8 (solid lines)
      and 2.0 (dashed--dotted lines), and 3 g/cm$^{3}$ (dashed lines) for orbital distances
      of 0.02, 0.05, and 0.1 AU (starting from the lower
      curve).}
         \label{single_nep}
   \end{figure}

\begin{figure}
   \centering
   \includegraphics[width=9cm]{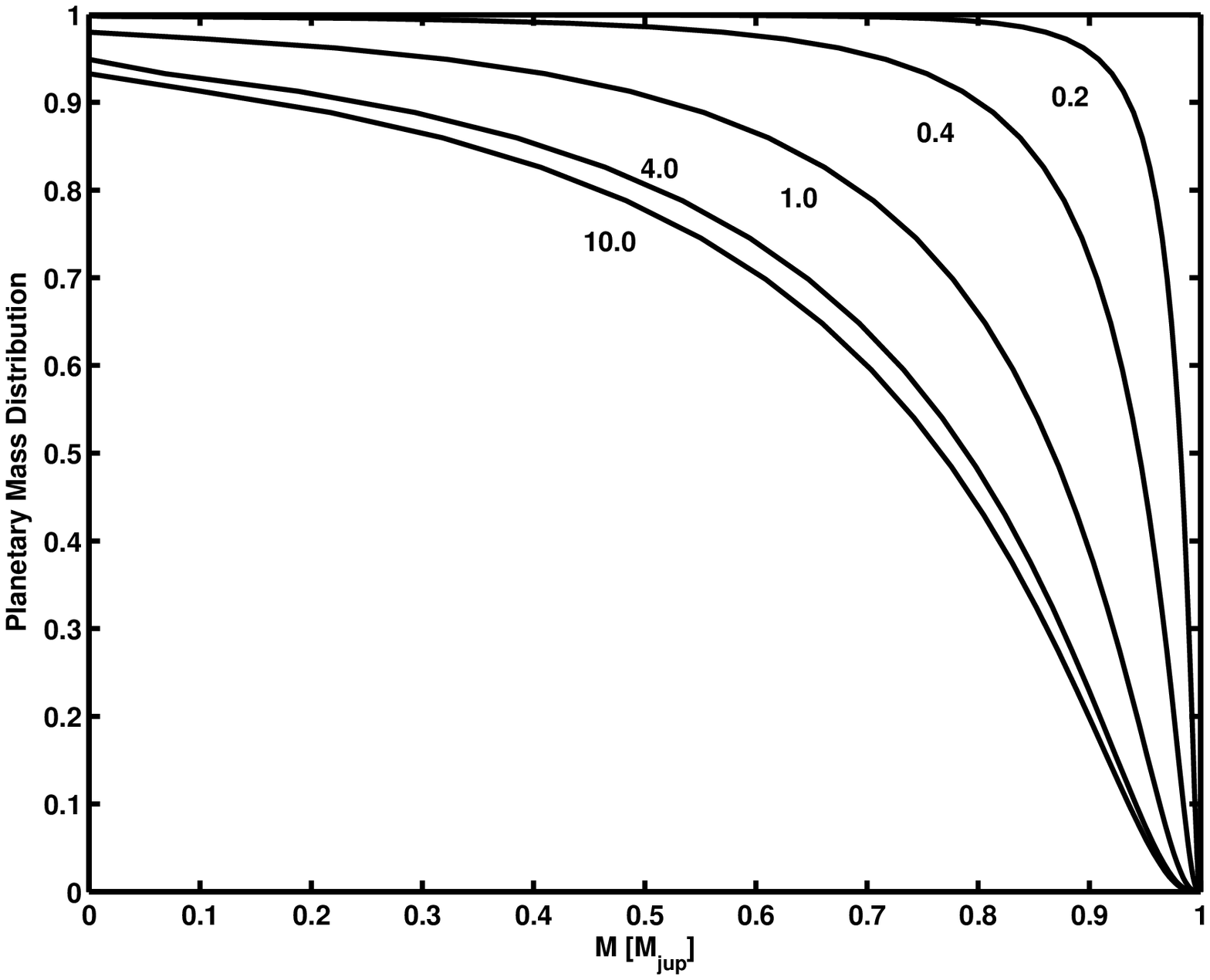}\hfill
   \includegraphics[width=9cm]{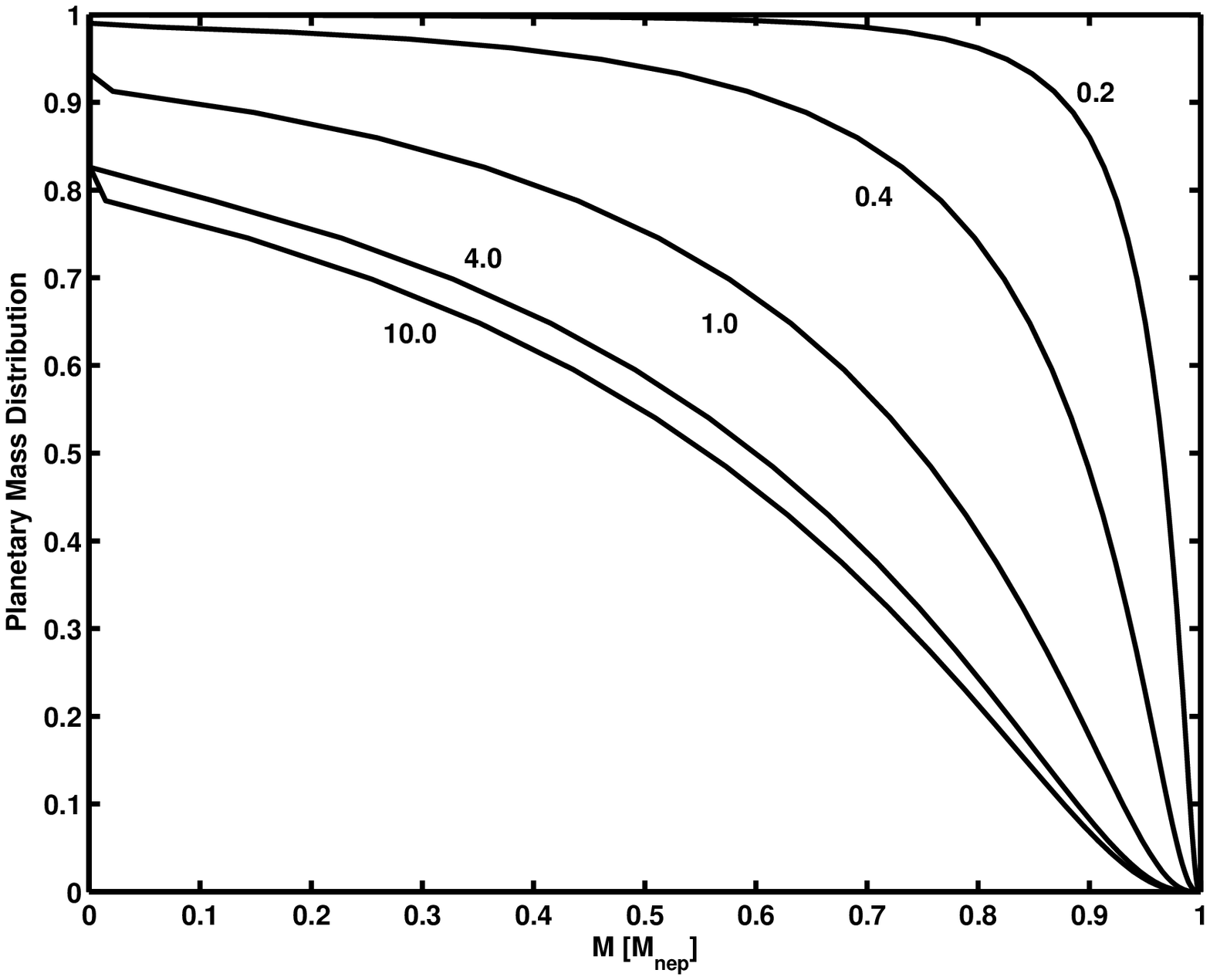}
      \caption{Mass distribution at 0.02 AU for a Jupiter--mass
      (upper panel, density is 0.4 g/cm$^{3}$) and a Neptune--mass planet
      (lower panel, density is 2 g/cm$^{3}$) at different ages of the system in Gyr.}
         \label{age}
   \end{figure}

If we use the luminosity distribution described above, it is
possible to calculate the mass lost from a planet for a given
density and orbital distance for a certain starting mass and a
given age of star. Because of lack of information, we assume that
the density of the planet remains constant in time. Some
implications of this assumption are discussed in Section 4. Thus,
we are able to determine the distribution of planetary mass
resulting from a given initial mass because of exposure to
different X--ray fluxes. It should be mentioned, that we are only
considering hydrogen--rich planets, meaning that we have no
information about a icy/rocky core, which can remain after
evaporating all the hydrogen. In Fig. \ref{single_jup}, the mass
distribution for an initial mass of 1 M$_{jup}$ and densities of
0.4 (corresponding to a low density planet like HD209458b) and 1.4
g/cm$^{3}$ (corresponding to a high density planet like TrES-2)
and orbital distances of 0.02, and 0.05 AU after 4 Gyr are shown.
For the closest orbit and low density, about 95 \% of the planets
can survive, and 50 \% have remaining masses of more than 0.8
M$_{jup}$. For a high density and an orbital distance of 0.02 AU,
more than 95 \% have a remaining mass of more than 0.8 M$_{jup}$,
while this number increases to nearly 100 \% for a larger orbital
distance of 0.05 AU. For orbital distance $\geq 0.05$ AU, only a
small fraction is affected by radiation coming from stars located
in the high--energy tail of the luminosity distribution. This is
in agreement with predictions by Hubbard et al. (2006, 2007) based
on a different approach that most of the Hot Jupiters are not
strongly influenced by mass loss
processes.\\
Fig. \ref{single_nep} shows the cumulative distribution function
for an initial mass of 1 M$_{nep}$ for the same orbital distances
but for densities of 0.8, 2 (corresponding to the density of the
transiting Hot Neptune observed by Gillon et al. (2007)), and 3
g/cm$^{3}$ after 4 Gyr. At a close--in orbit about 40 \% of
Neptune--sized planets with a density of 0.8 g/cm$^{3}$ can
survive, while this value increases to more than 80 \% for a
density of 2 g/cm$^{3}$, and to more than 90 \% for 3 g/cm$^{3}$.
At an orbital distance of 0.05 AU, less than 2 \% of the
low--density exoplanets would not survive the impact of their host
star radiation for 4 Gyr. Nearly 85 \% of the Neptune--mass
planets orbiting at 0.02 AU with a low density could be eroded to
Super--Earths (about 10 M$_\oplus$), and still more than 20 \% of
the high--density planets at the same orbits. Hydrogen envelopes
of 0.2 M$_{nep}$ can be easily lost at the close--in orbits. At
0.05 AU, 30 \% of the low--density planets can lose such an
envelope, while the number decreases to about 5 \% for
high--density planets. At 0.1 AU, also for Neptune--sized planets
the effects are very small.
Stopping our calculations after 4 Gyr is justified, which can be
seen from Fig. \ref{age}, where the mass distribution at 0.02 AU
for different ages of the system is shown for a Jupiter-- and a
Neptune--mass planet, respectively. One can see that the main loss
takes place in the first Gyr after the system's origin when high
energy stellar emission is higher. At later stages, the mass loss
is negligible compared with the loss during early stages.

\subsection{Distributions for an initial mass distribution}

\begin{figure}
   \centering
   \includegraphics[width=9cm]{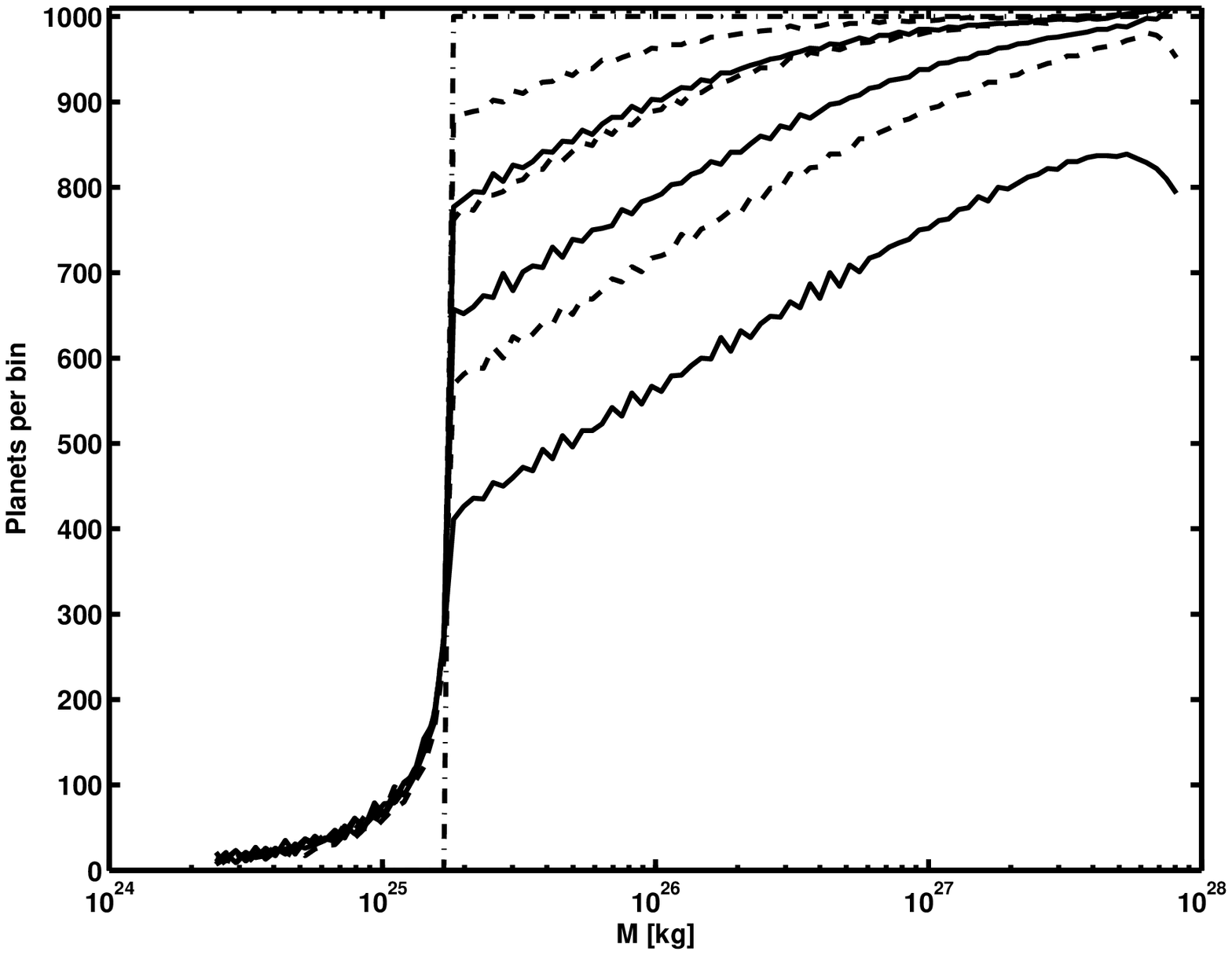}
      \caption{Initial flat mass distribution (dotted line)
      and final mass distributions after 4 Gyr for 0.4 (solid lines)
      and 1.4 g/cm$^{3}$ (dashed lines). Orbital distances
      are 0.02, 0.05, and 0.1 AU (starting from the lower
      curve). In the low--mass tail, the distribution is similar
      for all cases.}
         \label{dist}
   \end{figure}

After the investigation of the atmospheric loss for a single
planetary mass we proceed to evaluate the influence of loss
processes on an initial distribution of different mass. Since we
have no clear information on the initial mass distribution of
planets formed from a circumstellar disk, we assume a flat
distribution of masses ranging from 0.2 M$_{nep}$ up to 10
M$_{jup}$. Figure \ref{dist} shows the resulting distribution
after 4 Gyrs for different orbital distances and densities of 0.4
g/cm$^{3}$ and 1.4 g/cm$^{3}$. The percentage of planets lost for
0.02, 0.05, and 0.1 AU is 32 \%, 12 \%, and 4 \%, if we assume a
density of 0.4 g/cm$^{3}$. For the higher density, the
corresponding numbers are 18 \%, 5 \%, and 1 \%.
As expected, one can see that for the closest orbit and the lower
density, most planets are lost. Furthermore, the shape of the mass
distribution changes with a number of planets with masses smaller
than the minimum of the initial mass function. For distances
larger than 0.1 AU the initial and the final mass distribution are
similar because at these large distances the considered mass loss
processes do not affect the total mass of the planet anymore and
therefore do not alter the mass distribution.

\section{Discussion}

\begin{figure}
   \centering
   \includegraphics[width=9cm]{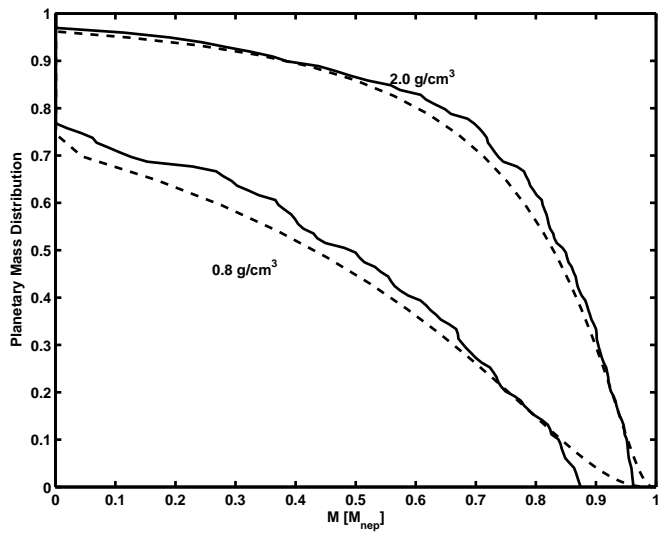}
      \caption{Planetary distribution function after 0.6 Gyr
      using a linear interpolation between the observed luminosity
      distribution functions of the Pleiades and the Hyades (solid lines)
      and using log--normal distribution functions (dashed lines) for
      two different densities and a Neptune--sized planet at 0.02 AU.}
         \label{comp_scal}
   \end{figure}

\begin{figure}
   \centering
   \includegraphics[width=9cm]{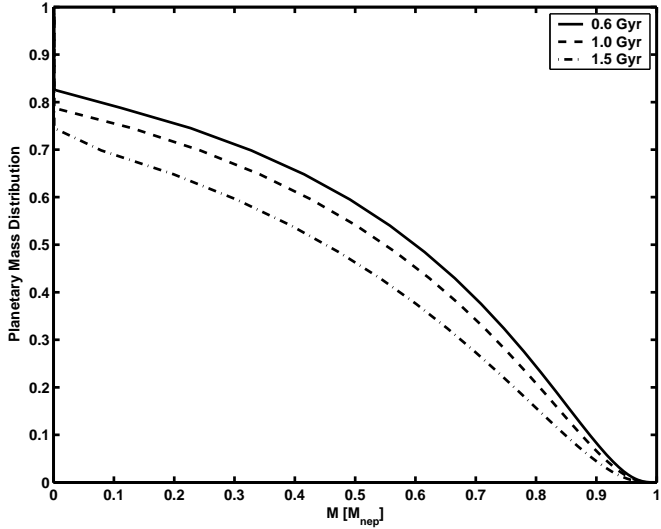}
      \caption{Planetary distribution function after 4.6 Gyr
      for different timing of the transition between the two
      scaling laws in Eq. \ref{scaling} for a Neptune--sized
      planet with a density of 2 g/cm$^{3}$ at 0.02 AU.}
         \label{shift_scal}
   \end{figure}

\begin{figure}
   \centering
   \includegraphics[width=9cm]{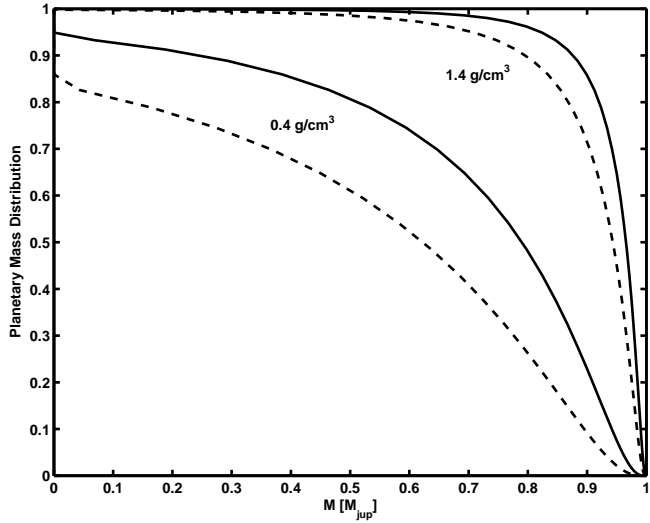}
      \caption{Planetary distribution function after 4.6 Gyr
      using a constant density (solid line) and for the scaling
      law derived by Lecavelier des Etangs (2007) (dashed lines) for a Jupiter--sized
      planet with a density of 0.4 and 1.4 g/cm$^{3}$ at 0.02 AU.}
         \label{den}
   \end{figure}

Since our work is subject to limited observational constraints,
several assumptions were used. In this section we discuss the
effects of some of these assumptions in particular for
Neptune--type planets where the effects are larger. Instead of the
observed luminosity distribution, an approximation by using
log--normal distributions was done. As can be seen in Fig.
\ref{comp}, the log--normal distributions fit the observed data
rather well. In order to check if the choice to use a log--normal
distribution function is justified, we additionally used the
observed data at 0.1 and 0.6 Gyr, for which we have reliable
observed luminosity functions, and applied a linear interpolation
in between and compared the results for the planetary distribution
function with the one achieved by using log--normal distribution
functions after 0.6 Gyr (Fig. \ref{comp_scal}).
One can see that even in the extreme case of a low--density planet
in a close--in orbit, the deviation is rather small. Only for
relatively unchanged masses (corresponding to the low--energy
tail) there are some differences since the log--normal
distribution is slightly different from the observed luminosity
distribution in this energy region. For the highly eroded planets
(corresponding to the high--energy tail) the agreement is very
good. For less extreme case, the curves are rather similar. Thus
we assume that the use of log--normal distribution functions
instead of the observed
values does not introduce significant errors.\\
Other uncertainties pertain to the scaling law used in Eq.
\ref{scaling}. The behavior of the X--ray luminosity over the
first 0.6 Gyrs is well known (Micela, 2002) but for older ages
there are no data available. It might be possible that the time
for the transition between the two scaling laws is not 0.6 Gyr but
at some older age. However, it cannot be older than 1.5 Gyr,
because for older ages we cannot reconstruct the observed
luminosity distribution function from Schmitt (1997) anymore. In
Fig. \ref{shift_scal} the effect of shifting the transition time
to older ages on the planetary mass distribution is shown.
For $t\leq 1$ Gyr, the difference is less than 5 \%, and even for
the rather unlikely case of $t$ = 1.5 Gyr the error is less than
10
\%.\\
In our study, we used the simplification that the density of the
planets remains constant over time. Lecavelier des Etangs (2007)
derived a simple scaling law for the radius of exoplanets with
$M_{p}\geq 0.1 M_{jup}$. Using this scaling law it is possible to
derive a scaling law for the temporal evolution of the density as
$1/(1+\beta t^{-0.3})^{3}$, where $\beta=0.2$ for $M_{p}\geq 0.3
M_{jup}$. We normalized this density to a density of 0.4 and 1.4
g/cm$^{3}$ for a planet with one Jupiter mass and an age of 4 Gyr
(Fig. \ref{den}). One can see that this effect has more impact
compared with the uncertainties discussed before. For a
low--density Hot Jupiter, it gives a difference of 10--20 \%
compared with the case of a constant density. The larger loss of
planets is caused by the effect that according to this scaling law
the initial density is smaller by about a factor of 2. As
expected, the effect is less pronounced for planets with higher
densities.\\
The derived mass function could be compared with the observed mass
distribution of known planets. However, we prefer to avoid to
compare our predictions with the present day planetary mass
distribution because of the complex biases present in the observed
sample. A main bias is due to the lack of active stars in the
observed sample. Indeed, active stars are eliminated from samples
since high activity of the star makes the detection of a planet
more difficult. On the other hand, the effect studied here is
particularly relevant for the most active stars with large X--ray
emissions. We expect that CoRoT will give us a sample of planet
without biases against activity for which it will be possible to
check our predictions.

\section{Conclusions}

We studied the influence of the stellar luminosity distribution on
atmospheric escape of exoplanets at orbits less than 0.1 AU. We
showed that a significant amount of planets can be evaporated over
time scales of 4 Gyr and that the final mass distribution is
different from the initial one. For a Jupiter--mass planet with a
density of 0.4 g/cm$^{3}$, about 5 \% of the planets are lost
after 4 Gyr at an orbit of 0.02 AU. For Neptune--mass planets this
number increases to about 60 \% for a density of 0.8 g/cm$^{3}$.
We also show that for close--in orbits a large number of planets
can be eroded to Super--Earth and/or may lose dense hydrogen
envelopes. Further, we present the resulting mass distribution of
an initially flat mass distribution of planets between 0.2
M$_{nep}$ and 10 M$_{jup}$. For close--in orbits about 32 \% of
the initial planets are lost and the distribution is
shifted to smaller masses.\\
Our work is a first step to understand the evolution of planetary
mass distribution due to the stellar activity evolution. Our
approach is subject to several assumptions that will be verified
with future observations and/or modelling. First of all we assume
a constant density in time and a heating function of 100 \%.
Better knowledge of the densities of exoplanets will be achieved
by further modelling supported by observational evidence, while
the latter assumption needs to be verified with detailed radiative
transfer calculations. The scaling law of the L$_{X}$ evolution in
Eq. \ref{scaling} is derived from a few points. This potential
source of uncertainty is however mitigated by the fact that most
of the effects occur in the first Gyr (see Fig. \ref{age}) where
the X--ray evolution is better known. The UV flux and evolution
should be added in the future. We plan to extend our work to lower
mass stars where the X--ray effects may influence the habitability
zone.

\begin{acknowledgements}
We thank the referee for useful comments, which helped to improved
the paper substantially. This work is supported by the Marie Curie
Fellowship Contract No. MTKD-CT-2004-002769 of the project ``The
influence of stellar high radiation on planetary atmospheres'',
and the host institution INAF--Osservatorio Astronomico di
Palermo. Also acknowledged is support by ASI--INAF contract
I/088/06/0.
\end{acknowledgements}

\end{document}